\documentclass[prb,superscriptaddress,twocolumn]{revtex4}
\usepackage{graphicx}

\begin{document}

\title{Charge-to-spin conversion of electron entanglement
states and
\\spin-interaction-free solid-state quantum computation}

\author{Wei-Min Zhang}
\email{wzhang@mail.ncku.edu.tw}
\affiliation{Department of Physics and Center for Quantum Information Science,
National Cheng Kung University, Tainan 70101, Taiwan}
\affiliation{National Center for Theoretical Science, Tainan 70101, Taiwan}

\author{Yin-Zhong Wu}
\affiliation{Department of Physics and Center for Quantum Information Science,
National Cheng Kung University, Tainan 70101, Taiwan}

\author{Chopin Soo}
\affiliation{Department of Physics and Center for Quantum Information Science,
National Cheng Kung University, Tainan 70101, Taiwan}
\affiliation{National Center for Theoretical Science, Tainan 70101, Taiwan}

\author{Mang Feng}
\affiliation{Department of Physics and Center for Quantum Information Science,
National Cheng Kung University, Tainan 70101, Taiwan}

\begin{abstract}

Without resorting to spin-spin coupling, we propose a scalable
spin quantum computing scheme assisted with a semiconductor
multiple-quantum-dot structure. The techniques of single electron
transitions and the nanostructure of quantum-dot cellular automata
(QCA) are used to generate charge entangled states of two
electrons, which are then converted into spin entanglement states
using single-spin rotations only. Deterministic two-qubit quantum
gates are also manipulated using only single-spin rotations with
the help of QCA. A single-shot readout of spin states can be
carried out by coupling the multiple dot structure to a quantum
point contact. As a result, deterministic spin-interaction-free
quantum computing can be implemented in semiconductor
nanostructure.

\end{abstract}



\maketitle

\section{Introduction}
Using electron spins to implement quantum information and quantum
computation in semiconductor nanostructure\cite{Loss98,Kane98}, as
one of the most important technology developments in spintronics,
has received tremendous attention in recent years. Prototypical
quantum computation schemes based on electron spins have been
proposed using gate voltage controlled \cite{Loss98,Div00}, and
optically driven \cite{Imam99,Sham02,Toiani03,Pazy03,Nazir04}
 spin-spin coupling in semiconductor quantum dots. However,
achieving a tunable spin-spin interaction with a sufficiently
large strength (compared to the strength of the charge Coulomb
interaction) is technically difficult. A spin-interaction-free
mechanism for logical operations on electron spins is therefore
more desirable. Many interaction-free schemes on measurement based
quantum computing have recently been proposed
\cite{Rau01,milburn,free,Engel05,Beige05,Barrett05}, but a robust,
deterministic, and scalable spin-interaction-free solid-state
quantum computing scheme in semiconductor nanostructure has yet to
emerge.

In this paper, we propose an implementation of scalable spin
quantum computation in semiconductor nanostructure without
resorting to spin-spin coupling. We can generate a charge
entangled state of two electrons using single electron
transitions\cite{RMP,PRset} assisted by a semiconductor
multiple-quantum-dot structure consisting of two double dots,
called the quantum-dot cellular automata
(QCA)\cite{lent93,science2}. The charge entangled state is then
converted into a spin entangled state using only single-spin
rotations. Spin-spin interaction is not required in this
implementation, and deterministic two-qubit controlled gates can
be easily manipulated as well. Therefore, deterministic and
scalable spin-interaction-free quantum computation can be
implemented in semiconductor nanostructure.

\section{architecture of the spin-interaction-free quantum computer}
The architecture of our scalable quantum computer is based on a
semiconductor multiple-quantum-dot structure schematically shown
in Fig.~1.
\begin{figure}[ht]
   \includegraphics[width=3.3in]{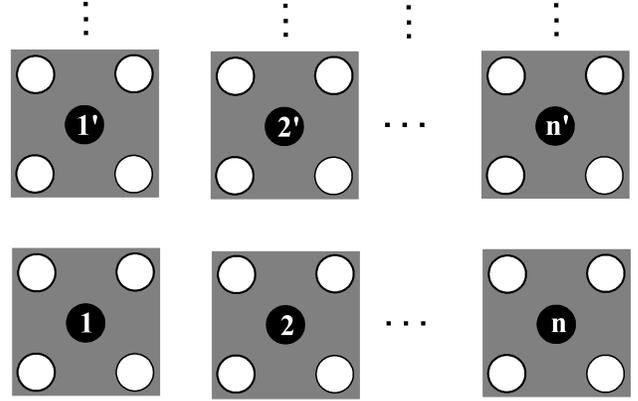}
   \caption{A schematic architecture of the scalable spin-interaction-free
   solid-state quantum computer
based on a multiple semiconductor quantum-dot structure. }
 \end{figure}
Each shaded squared box in Fig.~1 is regarded as a unit cell. Each
cell contains a qubit dot (the central black dot) surrounded by
four ancilla dots (the white dots). The detailed structure of a
unit cell is given in Fig.~2(a). The lines between quantum dots in
the cell indicate the possibility of interdot transitions. We
assume that each unit cell is charged with only one excess
conduction electron \cite{eninqd}. The electrostatic potential
energy ($\varepsilon_q$) of the excess electron in the qubit dot
is low enough compared to the energy ($\varepsilon_a$) in the
ancilla dots such that the electron sits initially in the qubit
dot due to the Coulomb blockade effect [see Fig.~2(b), where
$\varepsilon=\varepsilon_a-\varepsilon_q$]. Explicitly, we define
 quantum states of the excess electron sitting in the qubit dot
in each cell as a direct product of the spin-charge states, $|
S_{i} \rangle |e_{i}\rangle $ ($i=1,2, \cdots $), the
 charge states $|e_i \rangle$ are considered as ancilla
states, and the spin states $| S_{i}\rangle $ are chosen to be
qubit states in Pauli basis, $|\uparrow \rangle=|0 \rangle$ and
$|\downarrow \rangle =|1 \rangle$. A static uniform magnetic field
can be applied to split the qubit states $|0 \rangle$ and $|1
\rangle$ by the Zeeman energy for qubit initialization.

Furthermore, the four ancilla dots within a unit cell are coupled
to the qubit dot through gate voltages.  The electron in each cell
can be driven away from the qubit dot into ancilla dots only when
a two-qubit controlled operation is performed, and will be forced
to transit back as soon as the two-qubit operation has been
completed. Such transitions are controlled using gate voltage
pulses $V_i^{LR}, V_i^{TB}$. \cite{RMP,PRset,gorman} For instance,
by turning on the gate voltage $V_i^{LR}$, the electron will
transit to a certain site ($C_i$ or $D_i$) of the right two
ancilla dots [see Fig.~2(c)]. We denote the charge states of the
electron sitting in the ancilla dots A$_i, B_i, C_i$, and $D_i$ as
$|e_i^a\rangle$ with $a=A, B, C$ and $D$, respectively. The site
dependence of electron spin state is negligible in this
architecture. The effective Hamiltonian for the electron
transition between the qubit dot and ancilla dots can take the
form \cite{Fujikawa}
\begin{equation}
 H_i= \varepsilon(t)(|e_i^a\rangle\langle e_i^a|-
 |e_i\rangle \langle e_i|) + \Delta (|e_i^a\rangle\langle e_i|
 +|e_i\rangle \langle e_i^a|) , \label{ddh}
\end{equation}
where $\varepsilon(t)=\varepsilon-V_i^{^{LR}_{TB}}(t)$ and
$\Delta$ a tunneling coupling.
\begin{figure}[ht]
   \includegraphics[width=3.3in]{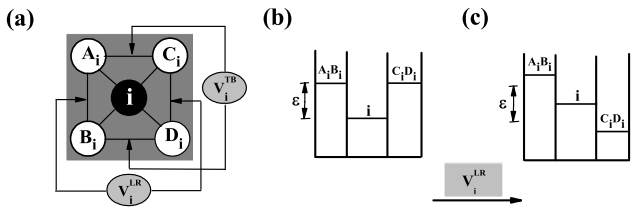}
   \caption{(a) The nanostructure of the unit cell $i$.
The gate voltages $V_i^{LR}$ and $V_i^{TB}$ control the electron
transitions among dots inside the cell, (b) and (c) show the
electrostatic potential energies of the electron at different dots
in the cell, without and with applying a gate voltage $V_i^{LR}$,
respectively.}
\end{figure}

Based on the above architecture, one can find that the two
double-dot pairs (e.g., $C_i$-$D_i$ and $A_j$-$B_j$ in Fig.~3)
between the qubit dots of two neighboring cells form a QCA. QCA
was originally proposed as a transistorless alternative to digital
circuit devices at nanoscale\cite{lent93,science2}. Recently,
semiconductor QCA has been fabricated from GaAs/AlGaAs
heterostructures\cite{SQCA} and from buried dopants\cite{BDCA} as
well.  Due to the Coulomb repulsion, when a QCA is charged with
two electrons, these two electrons will occupy one of the two
antipodal sites (called the charge polarization states denoted by
$|+\rangle =|e_i^De_j^A\rangle$ and $|- \rangle =
|e_i^Ce_j^B\rangle$, respectively). After a second-order
perturbation treatment, the effective Hamiltonian  of the QCA  is
then given by \cite{QCQCA}
\begin{equation} \label{eh-qda}
H_{\rm QCA}=(\omega+E_{\rm bias})P_{z} + \gamma P_{x},
\end{equation}
where $P_z \equiv {1\over 2}(|+\rangle\langle+| - |-\rangle
\langle -|)$ and $P_x \equiv {1\over 2}(|+\rangle\langle-| +
|-\rangle \langle +|)$, $\omega$ represents the energy offset of
the polarization states $|\pm \rangle$, coming from the on-site
electrostatic potential of each dot and the Coulomb repulsion
between dots, $E_{\rm bias}$ is an external bias polarization
applied to the QCA to adjust the energy splitting of the two
polarization states, and $\gamma$ accounts for the tunnelings
between the two polarization states controlled by gate voltages
acting on the two double-dot pairs. The electron tunneling between
different cells is forbidden by a built-in sufficiently high
energy barrier between the two neighboring cells.
\begin{figure}[ht]
   \includegraphics[width=3.1in]{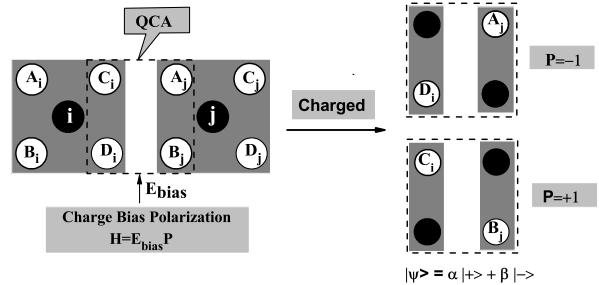}
   \caption{Quantum mechanically, the four quantum dots (the dotted
square boxes) between two qubit dots of the neighboring cells form
a coherent QCA.}
 \end{figure}

\section{charge-to-spin conversion of electron entanglement states}
A key to manipulate two-qubit controlled operations in this
architecture is the charge-to-spin conversion of two-electron
entanglement states. We shall use the QCA structure to generate a
charge entangled state via single electron transitions and then
convert it into a spin entangled state using single-spin rotations
only. Explicitly, consider a pair of neighboring unit cells,
e.g.~the $i$th and the $j$th cells (see Fig.~3). The initial state
of the two excess electrons is given as
\begin{equation} \label{initialstate}
|\Psi_0 \rangle = |S_ iS_j \rangle |e_i e_j \rangle .
\end{equation}
By turning on a positive voltage $V_i^{LR}$ and a negative voltage
$-V_j^{LR}$ to lower the on-site energy of the dots $C_i$-$D_i$
and $A_j$-$B_j$, the excess electrons in the two cells are
transited into the QCA consisting of the double-dot pairs
$C_i$-$D_i$ and $A_j$-$B_j$, and occupy one of the two polarized
states due to the Coulomb repulsion. To be specific, let the two
electrons occupy the polarized state
$|-\rangle=|e_i^Ce_j^B\rangle$, that is,
\begin{equation} \label{state2}
|\Psi_0 \rangle \stackrel{(V_i^{LR},-V_j^{LR})_{\rm
on}}{\longrightarrow} |\Psi_1 \rangle = |S_ iS_j \rangle |e_i^C
e_j^B \rangle .
\end{equation}
This manipulation is reliable in current experiments with passage
time of a few tens of picoseconds or less \cite{Fujikawa}. A
numerical simulation of electron transitions based on
Eq.~(\ref{ddh}) is presented in Fig.~4 (also refers
to\cite{setindqd}).
\begin{figure}[ht]
   \includegraphics[width=3.2in]{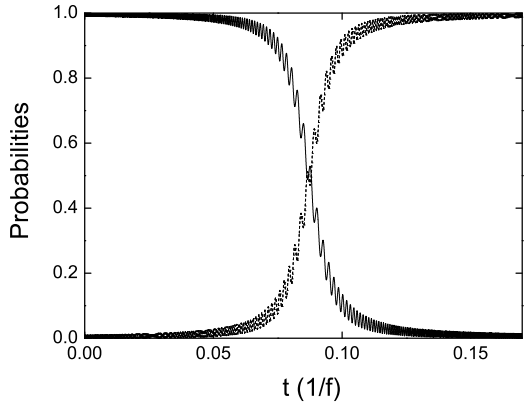}
   \caption{Numerical simulation of electron transition from a
   qubit dot to an ancilla dot based on Eq.~(\ref{ddh}), with
   $V_i^{LR}=A\sin(2\pi f t)$. The solid curve is the probability
   of electron sitting in the qubit dot during the transition
   and the dashed curve is that sitting in the ancilla dot.
   We take $\varepsilon:A:\Delta=1:2:0.1$ and
   $\varepsilon$ is of the order of meV. The transition
   can be completed in tens of picoseconds with a very high fidelity ($>$0.99).}
 \end{figure}

To generate a charge entangled state through the QCA structure, we
may adjust the external bias $E_{\rm bias}$ to make the two
polarized states degenerate (i.e., $E_{\rm bias}=-\omega $), and
then apply a ${\pi \over 2}$ gate voltage pulse (with a pulse time
$\tau_p={\pi\over 2}/\gamma$) to turn on the tunneling $\gamma$
between the two polarization states.
 The state $|\Psi_1\rangle$ thus becomes
\begin{eqnarray} \label{chargee}
|\Psi_2 \rangle &=& U_{\rm QCA}(\tau_p)|\Psi_1\rangle
   =e^{-iH_{\rm QCA}\tau_p} |\Psi_1\rangle
   \nonumber \\ &=&|S_ iS_j \rangle {1\over \sqrt{2}}
   \big(|e_i^C e_j^B \rangle - i |e_i^De_j^A\rangle
   \big).~~~~~
\end{eqnarray}
This superposition state is indeed a maximally entangled charge
state. Note that the current technique enables us to have nearly
identical dots for electronic tunneling \cite{Fujikawa,gorman}. As
long as no observable difference reflected in tunneling, the dots
in our design are not required to be fully identical. Meanwhile,
if the pulse duration $\tau_p$ (in picoseconds, see the discussion
later) can be accurate to femtoseconds, a very high fidelity of
Eq.~(\ref{chargee}), $F=\cos^2(\gamma \delta t)\sim 1-0.002$, is
obtainable.

Now, we shall convert the charge entangled state into a spin
entangled state using only single-spin rotations. Single-spin
manipulations in a single or double dots have been extensively
explored through voltage controls of a local magnetic field and
the local $g$ factor within nanoseconds \cite{slowspin} or using
ultrafast optical pulses up to picoseconds and
femtoseconds\cite{Kroutvar,Gupta}. Explicitly, consider the
initial spin state of the two electrons, $|S_i S_j
\rangle=|\uparrow \downarrow \rangle =|0 1\rangle$. Applying two
spin rotations on the electrons sitted at the dots $D_i$ and
$A_j$, respectively, $U_S^{DA}\equiv R_y^{A}(\pi)\otimes
R_x^{D}(\pi)$, where $R_k(\theta)\equiv \exp(-i\theta\sigma_k/2),
k=x,y,z$, the corresponding two-electron spin state at the dots
$D_i$ and $A_j$ becomes $U_S^{DA}|01 \rangle|e_i^De_j^A\rangle= -i
|1 0 \rangle|e_i^De_j^A\rangle$, while the spin state at the dots
$C_i$ and $B_j$ remains unchanged: $U_S^{DA}|01
\rangle|e_i^Ce_j^B\rangle= |01 \rangle|e_i^Ce_j^B\rangle$. After
the spin rotation operations, we turn off the gate voltages
$V_i^{LR}$ and $-V_j^{LR}$, the two electrons in the QCA are
transited back into the qubit dots $i$ and $j$, namely, the
electron charge states return to the initial states,
\begin{eqnarray} \label{bell1}
|\Psi_3 \rangle  = U_S^{DA}|\Psi_2 \rangle = \frac{1}{\sqrt{2}}
\big(|01 \rangle |e_i^C e_j^B \rangle -|10 \rangle|e_i^D e_j^A
\rangle \big)
\nonumber \\
\stackrel{(V_i^{LR},-V_j^{LR})_{\rm off}}{\longrightarrow}
 \frac{1}{\sqrt{2}}\big(|0 1 \rangle  -  |10 \rangle \big)|e_i e_j
 \rangle .
\end{eqnarray}
As a result, the electron charge entangled state has been
converted completely into the spin entangled state, the Bell state
$|\psi^-\rangle$. This is the implementation of charge-to-spin
conversion of a two-electron entanglement state. Repeating the
process of Eqs.~(\ref{initialstate}-\ref{bell1}) with different
initial spin states $|10 \rangle,|00 \rangle$, and $|11 \rangle $,
we obtain
\begin{equation} \label{bell2}
\overbrace{U_S^{DA} U_{\rm QCA}(\tau_p)}^{(V_i^{LR},-V_j^{LR})}
\left\{
\begin{array}{c}
|1 0 \rangle |e_i e_j \rangle \\
|0 0 \rangle |e_i e_j \rangle \\
|1 1 \rangle |e_i e_j \rangle \\
\end{array}\right.
=
\begin{array}{c}
{1\over\sqrt{2}}\big(|10\rangle+|01\rangle\big)|e_i e_j
\rangle \\
{1\over\sqrt{2}}\big(|00\rangle-|11\rangle\big)|e_i e_j
\rangle \\
{1\over\sqrt{2}}\big(|11\rangle+|00\rangle\big)|e_i e_j \rangle,
\end{array}
\end{equation}
namely, the other three spin Bell states ($|\psi^+\rangle,
|\phi^-\rangle, |\phi^+\rangle$) can be generated from the same
quantum operations.

\section{The spin-interaction-free quantum computing and decoherence analysis}
Once we can convert charge entangled states into spin entangled
states, manipulating a two-qubit gate for electron spin qubits is
relatively simple. Consider the CNOT gate as an example. Instead
of using the spin rotation $U_S^{DA}$ in Eq.~(\ref{bell1}), we
should apply a single spin rotation on each dot $C_i$, $D_i$,
$A_j$, and $B_j$ in the QCA with the rotation operator
$U_S^{CDAB}= R_x^B(\frac{\pi}{2})\otimes
R_x^A(\frac{3\pi}{2})\otimes R_z^D(\frac{3\pi}{2})\otimes
R_z^C(\frac{\pi}{2})$ to rotate the corresponding electron spin
state in $|\Psi_2\rangle$. Using the same process of
Eqs.~(\ref{initialstate}-\ref{bell1}) with the above replacement
of the spin rotations, we have
\begin{equation} \label{cnot}
\overbrace{U_S^{CDAB} U_{\rm QCA}(\tau_p)}^{(V_i^{LR},-V_j^{LR})}
|\Psi_0\rangle = U_{\rm CNOT} |\Psi_0 \rangle .
\end{equation}
It is easy to check that Eq.~(\ref{cnot}) gives explicitly the
CNOT gating: $|0 0 \rangle |e_i e_j \rangle \rightarrow |0 0
\rangle |e_i e_j \rangle$, $|0 1 \rangle |e_i e_j
\rangle\rightarrow |0 1 \rangle |e_i e_j \rangle$, $|1 0 \rangle
|e_i e_j \rangle\rightarrow |1 1 \rangle |e_i e_j \rangle$, and
$|1 1 \rangle |e_i e_j \rangle \rightarrow |1 0 \rangle |e_i e_j
\rangle$. Thus, a spin two-qubit CNOT gate is manipulated using
only single-spin rotations and single electron transitions
assisted with the QCA structure. Combining the spin two-qubit CNOT
gate with single spin rotations, universal quantum computation has
been achieved without resorting to spin-spin coupling for the
first time.

Meantime, the qubit state readout in this scheme can be realized
as follows. A single-shot readout of electron spin states in qubit
dot can be realized by coupling the unit cell to a quantum point
contact (QPC). Explicitly, one can tune a gate voltage pulse,
e.g., $V_i^{LR}$, to lower the on-site energy of dots $C_i,$ and
$D_i$ such that the electron will remain in the qubit dot if it is
in the state $|0\rangle$; otherwise, the electron will transit to
the dots $C_i$ and $D_i$ and then return to the qubit dot after
the pulse if it is in the state $|1\rangle$. Thus, one can
read-out qubit states by measuring charge current changes through
the QPC channel. Such a measurement has been experimentally
demonstrated in semiconductor dots\cite{spinm} and theoretically
studied extensively\cite{Gurvitz97,lee06}.

The possible imperfection in the above manipulation may come from
 decoherence and inaccurate operations on electron states.  The
effect of gate voltage pulses on the electrically floating double
dots involves an abrupt change in the confinement potential and
modifies $\Delta, \omega$ as well as $\varepsilon, \gamma$ in
Eqs.~({\ref{ddh}-\ref{eh-qda}). Combining with the noises from the
electron-phonon interaction and piezoelectric coupling results in
a typical charge decoherence time $T_2 \sim 1-10$ ns in GaAs dots
\cite{Fujikawa}. The decoherence could not affect the transition
of Eq.~(\ref{state2}) since the tunneling only occurs near the
resonance region $\varepsilon(t) \sim 0$ in picoseconds (see Fig.
4). For the operation in (\ref{chargee}), the tunneling coupling
inside the QCA should be controlled with the requirement,
$Ke^{2}/r < \gamma < Ke^{2}/d$, where $K=6.9\times 10^{8} N\cdot
m^{2}/C^{2}$ for GaAs, $e$ is the electron charge, and $r=\sqrt
{2} d$ with $d$ the spacing between $A_{j}$ ($C_{i}$) and $B_{j}$
($D_{i}$). Direct calculation for $d \sim 50- 100 $ nm shows that
$\gamma$ is of the order of terahertz. Accordingly the operation
in (\ref{chargee}) could also be done within picoseconds, much
shorter than the charge decoherence time in double dots.

Since the charge decoherence time is only a few nanoseconds short,
a very fast and elaborately operated single-spin rotation is
required for single spin rotations in Eqs.~(\ref{bell1}) and
(\ref{cnot}). Recent experiments demonstrated that optical tipping
pulses with a frequency below the band gap of the semiconductor
nanostructure can create an effective magnetic field in the order
of 20 T via the optical Stark effect, which can induce substantial
rotations of electron spins at the femtosecond scale\cite{Gupta}.
Meanwhile, spin-flip Raman transitions using the adiabatic process
of two ultrafast laser pulses\cite{Kroutvar} can also fully
control single-spin rotations in semiconductor quantum dots at
picosecond or femtosecond scale\cite{Chen04}. These optical
controls of single spin rotations are technically very supportive
for a practical implementation of the present scheme.

As an overall decoherence analysis, recent measurements in GaAs
and In(Ga)As quantum dots have shown a long spin relaxation time
($T_1 \sim 1-20$ ms)\cite{spinT1}, the spin decoherence time
($T_2$) in GaAs dots caused by the complicated nuclear spin
fluctuation is about $10$ ns \cite{spinT2,Patta}. A lower bound on
the spin coherence time exceeding 1 ms has also been established
using spin-echo techniques on two electrons in double
dots\cite{Patta}, while the charge decoherence time is a few
ns\cite{Fujikawa} and maybe up to 200 nanoseconds in isolated
silicon double dots\cite {gorman}. If the manipulation of single
electron transitions and single-spin rotations can be completed
within picoseconds to femtoseconds, the implementation of
spin-interaction-free quantum computation with quantum error
correction is reliable in experiments.

\section{summary}
In summary, without resorting to spin-spin coupling, we have
proposed a deterministic and scalable spin quantum computing
scheme assisted by a semiconductor multiple-quantum-dot structure.
Spin-interaction-free solid-state quantum computing is a big
challenge, in principle. In this scheme, we are able to achieve
such an implementation basically relying on the charge-to-spin
conversion of electron entanglement states with the help of the
QCA. The QCA structure offers an intrinsic charge coupling of two
electrons, which is more effective, completely deterministic, and
scalable in comparison with the measurement based quantum
computing scheme in semiconductor nanostructure\cite{Engel05}.
Since spin couplings are much weaker than the charge Coulomb
interaction, such a spin-interaction-free quantum computing has
the advantage of being robust against the technical difficulties
of electronically or optically generating tunable spin-spin
couplings\cite{Loss98,Div00,Imam99,Sham02,Toiani03,Pazy03,Nazir04}.
The present scheme only involves gate voltage controls of electron
transitions and the optical manipulation of spin coherence in
semiconductor dots. These techniques are currently reliable in
experiments. Therefore, the spin-interaction-free quantum
computing can be realized practically in semiconductor
nanostructure.

\begin{acknowledgments}
This work is supported by the National Science Council of Taiwan,
Republic of China under Contracts No.~NSC-94-2120-M-006-003,
No.~NSC-94-2112-M-006-007, and No.~NSC-95-2112-M-006-001.
\end{acknowledgments}

%\newpage


\begin{thebibliography}{99}

\bibitem{Loss98}D. Loss and D. P. DiVincenzo,
Phys. Rev. A {\bf 57}, 120 (1998).

\bibitem{Kane98}B. E.~Kane, Nature (London) {\bf 393}, 133 (1998).

\bibitem{Div00}G. Burkard, D. Loss, and D. P. DiVincenzo, Phys. Rev. B {\bf 59}, 2070 (1999);
D. P. DiVincenzo, D. Bacon, J. Kempe, G. Burkard, and K. B.
Whaley, Nature (London) {\bf 408}, 339 (2000).

\bibitem{Imam99} A. Imamoglu, D. D. Awschalom, G. Burkard, D. P. DiVincenzo,
D. Loss, M. Sherwin, and A. Small, Phys. Rev. Lett. {\bf 83}, 4204
(1999);

\bibitem{Sham02}C. Piermarocchi, P. Chen, L. J. Sham, and D. G. Steel, Phys. Rev. Lett. {\bf 89}, 167402 (2002).

\bibitem{Toiani03} F. Troiani, E. Molinari, and U. Hohenester, Phys. Rev. Lett.
{\bf 90}, 206802 (2003).

\bibitem{Pazy03} E. Pazy, E. Biolatti,T. Calarco , I. D'Amico, P. Zanardi,
F. Rossi and P. Zoller, Europhys. Lett. {\bf 62} 175 (2003); M.
Feng, I. D'Amico, P. Zanardi, and F. Rossi, Europhys. Lett. {\bf
66}, 14 (2004).

\bibitem{Nazir04} A. Nazir,  B. W. Lovett, S. D. Barrett, T. P. Spiller,
and G. A. D. Briggs, Phys. Rev. Lett. {\bf 93}, 150502 (2004);
B.~W. Lovett, A. Nazir, E. Pazy, S. D. Barrett, T. P. Spiller, and
G. A. D. Briggs, Phys. Rev. B {\bf 72}, 115324 (2005).

\bibitem{Rau01} R. Raussendorf and H. J. Briegel, Phys. Rev. Lett. {\bf 86}, 5188
(2001).

\bibitem{milburn}E. Knill, R. Laflamme, and G. J. Milburn,
Nature {\bf 409}, 46 (2001).

\bibitem{free}C. W. J. Beenakker, D. P. DiVincenzo, C. Emary, and
M. Kindermann, Phys. Rev. Lett. {\bf 93}, 020501 (2004).

\bibitem{Engel05} H. A. Engel and D. Loss, Science {\bf 309}, 586
(2005).

\bibitem{Beige05} Y. L. Lim, A. Beige, and L. C. Kwek, Phys. Rev.
Lett. {\bf 95}, 030505 (2005).

\bibitem{Barrett05} S. D. Barrett and P. Kok, Phys. Rev. A {\bf 71}, 060310(R)
(2005).

%\bibitem{no-go}B.~M.~Terhal, and D.~P.~DiVincenzo, Phys. Rev. A
%{\bf 65}, 032325 (2002); E.~Knill, quant-ph/0108033.

\bibitem{RMP} W. G. van der Wiel, S. De Franceschi, J. M. Elzerman,
T. Fujisawa, S. Tarucha, and L. P. Kouwenhoven, Rev. Mod. Phys. {\bf 75},
1 (2003)

\bibitem{PRset} T. Fujisawa, T. Hayashi, and S. Sasaki, Rep. Prog. Phys.
{\bf 69}, 759 (2006).

\bibitem{lent93} C. S. Lent, P. D. Tougaw, W. Porod, and G. H. Bernstein,
Nanotechnology {\bf 4}, 49 (1993); P. D. Tougaw and C. S. Lent, J.
Appl. Phys. {\bf 75}, 1818 (1994).

\bibitem{science2}A. O. Orlov, I. Amlani, G. H. Bernstein, C. S. Lent,
and G. L. Snider, Science {\bf 277}, 928 (1997); I. Amlani, A. O.
Orlov, G. Toth, G. H. Bernstein, C. S. Lent, G. L. Snider, {\it
ibid.} {\bf 284}, 289(1999).

\bibitem{eninqd} M. Ciorga, A.S. Sachrajda, P. Hawrylak, C. Gould,
P. Zawadzki, S. Jullian, Y. Feng, and Z. Wasilewski, Phys. Rev. B
{\bf 61}, R16315 (2000); J. M. Elzerman, R. Hanson, J. S.
Greidanus, L. H. Willems van Beveren, S. D. Franceschi, L. M. K.
Vandersypen, S. Tarucha, and L. P. Kouwenhoven, {\it ibid.} {\bf
67}, 161308 (2003).

\bibitem {gorman} J. Gorman, D. G. Hasko, and D. A. Williams,
Phys. Rev. Lett. {\bf 95}, 090502 (2005).

\bibitem{Fujikawa}T.~Hayashi, T. Fujisawa, H. D. Cheong, Y. H. Jeong,
and Y. Hirayama, Phys. Rev. Lett. {\bf 91}, 226804 (2003).

\bibitem{SQCA}S. Gardelis, C. G. Smith, J. Cooper, D. A. Ritchie,
E. H. Linfield, and Y. Jin, Phys. Rev. B {\bf
67}, 033302 (2003).

\bibitem{BDCA}J. H. Cole, A. D. Greentree, C. J. Wellard,
L. C. L. Hollenberg, and S. Prawer, Phys. Rev. B {\bf 71}, 115302
(2005).

\bibitem{QCQCA}G. Toth, and C. S. Lent, Phys. Rev. A
{\bf 63}, 052315 (2001).

\bibitem{setindqd}M. Forre, J. P. Hansen, V. Popsueva, and
A. Dubois, Phys. Rev. B {\bf 74}, 165304 (2006).

\bibitem{slowspin}M. Xiao, I. Martin, E. Yablonovitch and
H. W. Jiang, Nature {\bf 430}, 435 (2004); Y. Kato, R. C. Myers,
D. C. Driscoll, A. C. Gossard, J. Levy, D. D. Awschalom, Science
{\bf 299}, 1201 (2003).

\bibitem{Kroutvar} N. V. Vitanov, T. Halfmann, B.W. Shore, and K. Bergmann,
Annu. Rev. Phys. Chem. {\bf 52}, 763 (2001).

\bibitem{Gupta} J. A. Gupta, R. Knobel, N. Samarth, D. D. Awschalom,
Science {\bf 292}, 2458 (2001).

\bibitem{spinm}J.M. Elzerman, R. Hanson, L. H. Willems van Beveren,
B. Witkamp, L. M. K. Vandersypen and L. P. Kouwenhoven, Nature
{\bf 430}, 431 (2004); R. Hanson, B. Witkamp, L. M. K.
Vandersypen, L. H. W. van Beveren, J. M. Elzerman, and L. P.
Kouwenhoven, Phys. Rev. Lett. {\bf 94}, 196802 (2005).

\bibitem{Gurvitz97} S. A. Gurvitz, Phys. Rev. B {\bf 56}, 15215 (1997).

\bibitem{lee06} M.T. Lee and W.M. Zhang, Phys. Rev. B \textbf{74}, 085325
(2006).

\bibitem{Chen04}P.~Chen, C. Piermarocchi, L. J. Sham, D. Gammon,
and D. G. Steel, Phys. Rev. B {\bf 69}, 075320 (2004).

\bibitem{spinT1} T. Fujisawa, D. G. Austing, Y. Tokura, Y. Hirayama,
and S. Tarucha, Nature 419, 278 (2002); M. Kroutvar, Y. Ducommun,
D. Heiss, M. Bichler, D. Schuh, G. Abstreiter and J. J. Finley,
{\it ibid.} {\bf 432}, 81 (2004).

\bibitem{spinT2} F.H.L. Koppens, J. A. Folk, J. M. Elzerman, R. Hanson,
L. H. W. van Beveren, I. T. Vink, H. P. Tranitz, W. Wegscheider,
L. P. Kouwenhoven, and L. M. K. Vandersypen,  Science {\bf 309},
1346 (2005);  M.V. Gurudev Dutt, J. Cheng, B. Li, X. Xu, X. Li, P.
R. Berman,1 D. G. Steel, A. S. Bracker, D. Gammon, S. E. Economou,
R. B. Liu, and L. J. Sham, Phys. Rev. Lett. {\bf 94}, 227403
(2005).

\bibitem{Patta} J. R. Petta, A. C. Johnson, J. M. Taylor, E. A. Laird,
A. Yacoby, M. D. Lukin, C. M. Marcus, M. P. Hanson, and A. C.
Gossard, Science {\bf 309}, 2180 (2005).

\end{thebibliography}
\end{document}